\documentclass[dvips]{acta}
\usepackage{supertabular,lscape,epsfig}
\usepackage{amssymb}
\usepackage{amsmath}
\DeclareSymbolFont{ppa}{OT1}{ppl}{m}{it}
\DeclareMathSymbol{\vv}{\mathalpha}{ppa}{'166}

\newfont{\hb}{rphvb at 10pt}%bezszeryfowe pó³grube
\newfont{\hbo}{rphvbo at 10pt}%bezszeryfowe pó³grube kursywa
\newfont{\bitt}{rptmbi at 12pt}%pó³gruba kursywa (tytu³ artyku³u)
\newfont{\bits}{rptmbi at 11pt}%pó³gruba kursywa (tytu³y rozdzia³ów)

\SetPages{153}{162}

\SetVol{58}{2008}

\begin{document}

%Zwarte naglowki, jeden wiersz
\newcommand{\TabCapp}[2]{\begin{center}\parbox[t]{#1}{\centerline{
  \small {\spaceskip 2pt plus 1pt minus 1pt T a b l e}
  \refstepcounter{table}\thetable}
  \vskip2mm
  \centerline{\footnotesize #2}}
  \vskip3mm
\end{center}}

%Zwarte naglowki, dwa wiersze
\newcommand{\TTabCap}[3]{\begin{center}\parbox[t]{#1}{\centerline{
  \small {\spaceskip 2pt plus 1pt minus 1pt T a b l e}
  \refstepcounter{table}\thetable}
  \vskip2mm
  \centerline{\footnotesize #2}
  \centerline{\footnotesize #3}}
  \vskip1mm
\end{center}}

%Zwarte naglowki, jeden wiersz
\newcommand{\MakeTableSepp}[4]{\begin{table}[p]\TabCapp{#2}{#3}
  \begin{center} \TableFont \begin{tabular}{#1} #4 
  \end{tabular}\end{center}\end{table}}

%Zwarte naglowki, jeden wiersz
\newcommand{\MakeTableee}[4]{\begin{table}[htb]\TabCapp{#2}{#3}
  \begin{center} \TableFont \begin{tabular}{#1} #4
  \end{tabular}\end{center}\end{table}}

%Zwarte naglowki, dwa wiersze
\newcommand{\MakeTablee}[5]{\begin{table}[htb]\TTabCap{#2}{#3}{#4}
  \begin{center} \TableFont \begin{tabular}{#1} #5 
  \end{tabular}\end{center}\end{table}}

%FWHM, PSF - proste, MgII, H$\alpha$
\newfont{\bb}{ptmbi8t at 12pt}
\newfont{\bbb}{cmbxti10}
\newfont{\bbbb}{cmbxti10 at 9pt}
\newcommand{\uprule}{\rule{0pt}{2.5ex}}
\newcommand{\douprule}{\rule[-2ex]{0pt}{4.5ex}}
\newcommand{\dorule}{\rule[-2ex]{0pt}{2ex}}
\def\thefootnote{\fnsymbol{footnote}}
\begin{Titlepage}
\Title{The Optical Gravitational Lensing Experiment.\\
Triple-Mode and 1O/3O Double-Mode Cepheids
in the Large Magellanic Cloud\footnote{Based on observations obtained
with the 1.3-m Warsaw telescope at the Las Campanas Observatory of the
Carnegie Institution of Washington.}}
\Author{I.~~S~o~s~z~y~\'n~s~k~i$^1$,~~
R.~~P~o~l~e~s~k~i$^1$,~~
A.~~U~d~a~l~s~k~i$^1$,~~
M.~~K~u~b~i~a~k$^1$,\\
M.\,K.~~S~z~y~m~a~{\'n}~s~k~i$^1$,~~
G.~~P~i~e~t~r~z~y~\'n~s~k~i$^{1,2}$,~~
\L.~~W~y~r~z~y~k~o~w~s~k~i$^{1,3}$,\\
O.~~S~z~e~w~c~z~y~k$^{1,2}$~~
and~~ K.~~U~l~a~c~z~y~k$^1$}
{$^1$Warsaw University Observatory, Al.~Ujazdowskie~4, 00-478~Warszawa, Poland\\
e-mail:
(soszynsk,rpoleski,udalski,mk,msz,pietrzyn,wyrzykow,szewczyk,kulaczyk)@astrouw.edu.pl\\
$^2$ Universidad de Concepci{\'o}n, Departamento de Fisica, Casilla 160--C,
Concepci{\'o}n, Chile\\
$^3$ Institute of Astronomy, University of Cambridge, Madingley Road,
Cambridge CB3 0HA, UK}
\Received{July 25, 2008}
\end{Titlepage}

\Abstract{We report the discovery of three new triple-mode classical 
Cepheids in the Large Magellanic Cloud, two of them with the fundamental,
first overtone and second overtone excited, and one pulsating
simultaneously in the first three overtones. Thus, the number of
triple-mode Cepheids in the LMC is increased to five. We also present two
objects belonging probably to a new type of double-mode Cepheids having the
first and third overtones excited. We measure the rates of period change in
these stars and detect decrease of periods in two of them, what is in
conflict with theoretical predictions.}{Cepheids -- Stars: oscillations --
Stars: evolution -- Magellanic Clouds}

\Section{Introduction}
Classical Cepheids pulsating simultaneously in three radial modes are
extremely rare objects. Only four triple-mode Cepheids and $\delta$~Sct
stars are known in the Galaxy: AC~And (Fitch and Szeidl 1976), V823~Cas
(Antipin 1997), V829~Aql (Handler \etal 1998) and recently discovered GSC
762-110 (Wils \etal 2008). All these stars have fundamental (F), first (1O)
and second (2O) overtone modes excited. The first two objects have
fundamental-mode periods around 0.7~days and can be recognized as classical
Cepheids (though some controversy about this classifications exists), while
two other triple-mode pulsators can be classified as $\delta$~Sct stars,
because their longest periods are below 0.3~days.

Moskalik \etal (2004) used the OGLE-II photometry (Soszy{\'n}ski \etal
2000) to discover two Cepheids in the Large Magellanic Cloud (LMC)
pulsating simultaneously in the first, second and third overtones. These
stars were theoretically analyzed by Moskalik and Dziembowski (2005), who
concluded that they must be on the first crossing of the instability
strip. Triple-mode pulsators are very useful in studies of stellar
interiors, because three periods strictly constrain the stellar parameters
(\eg Kov{\'a}cs and Buchler 1994, Moskalik and Dziembowski 2005, Jurcsik
\etal 2006). Periods are those observables, that can be measured with very
high precision, independently of the interstellar extinction and
photometric calibration.

Here, we report the discovery of three triple-mode Cepheids in the LMC, two
of which pulsate in the fundamental mode, first and second overtones, and
one of the same type as discovered by Moskalik \etal (2004), \ie with the
first three overtones excited. We also show a new class of double mode
Cepheids -- stars oscillating simultaneously in the first and the third
overtones.

\begin{figure}[htb]
\includegraphics[width=12.5cm, bb=40 225 565 745]{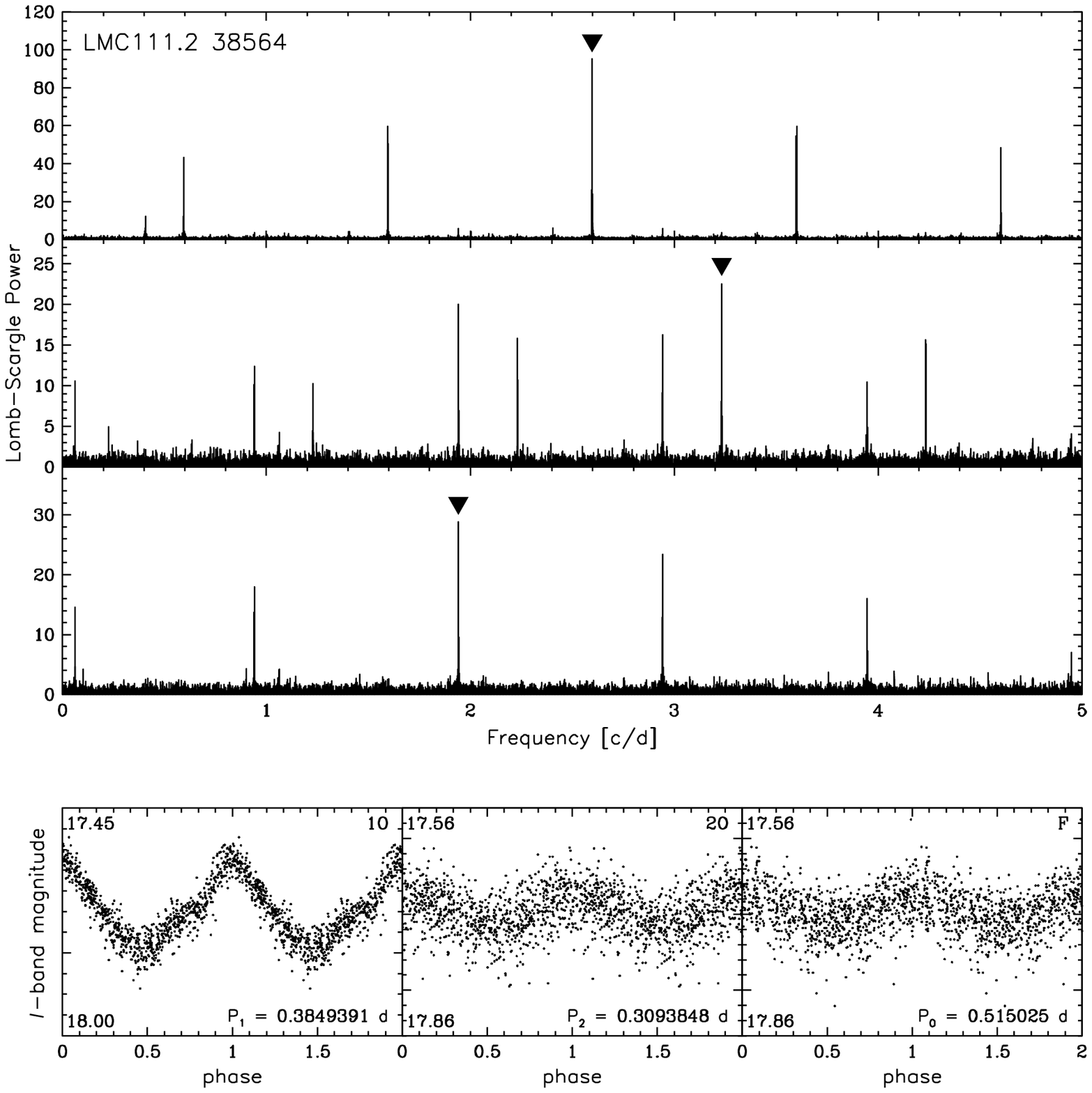}
\FigCap{Power spectra and light curves of F/1O/2O triple-mode Cepheid 
LMC111.2 38564. {\it Upper panel} displays periodogram obtained for the
original data, {\it middle panel} shows data after subtracting the primary
period marked with triangle, and the {\it third panel} displays the power
spectrum after prewhitening with the primary and the secondary
periods. Three {\it bottom panels} show light curves folded with the
primary, secondary and tertiary periods after removing the other two
modes.}
\end{figure}

\Section{Observational Data}
The photometry used in this study was obtained in the course of the third
phase of the Optical Gravitational Lensing Experiment (OGLE-III) with the
Warsaw Telescope at Las Campanas Observatory, Chile (operated by Carnegie
Institution of Washington). Typically about 380 Cousin's {\it I}-band
observations per star were collected between 2001 and 2008 for 116 fields
covering about 40 square degrees in the LMC. For three objects (LMC111.2
38564, LMC161.6 14207 and LMC169.8 75460) the OGLE-II data were available,
what increased the number of points to 800--900 per star, collected within
12 years. In addition a few dozen observations in Johnson's $V$ waveband
were done to obtain the $(V-I)$ color information. Detailed description of
the data reduction, photometric calibration and astrometric transformations
of the OGLE-III data can be found in Udalski \etal (2008).

\Section{Search for Multiperiodic Cepheids}
We searched for multiperiodicity in our sample of more than 3300 classical
Cepheids in the LMC (Soszy{\'n}ski \etal 2008) using standard
procedure. Each light curve was folded with the primary pulsational period
and a Fourier series of an order depending on the shape and scatter of the
light curve was fitted. The number of harmonics of the Fourier
decomposition was adjusted to minimize the $\chi^2$ per one degree of
freedom. Then, the fitted function was subtracted from the light curve and
the period searching was carried out on the residual data. This procedure
was repeated twice.

We found that way a considerable number of 264 double-mode Cepheids of both
types: F/1O and 1O/2O. Among bimodal variables we also found interesting
objects with ratios of periods inconsistent with well known values for
double-mode Cepheids, for example significant number of stars with period
ratios close to 1 (corresponding to non-radial pulsations), or in the range
0.60--0.63 (of unknown origin). All these objects have been published in the
first part of the OGLE-III Catalog of Variable Stars presenting classical
Cepheids in the LMC.

\begin{figure}[htb]
\includegraphics[width=12.5cm, bb=40 225 565 745]{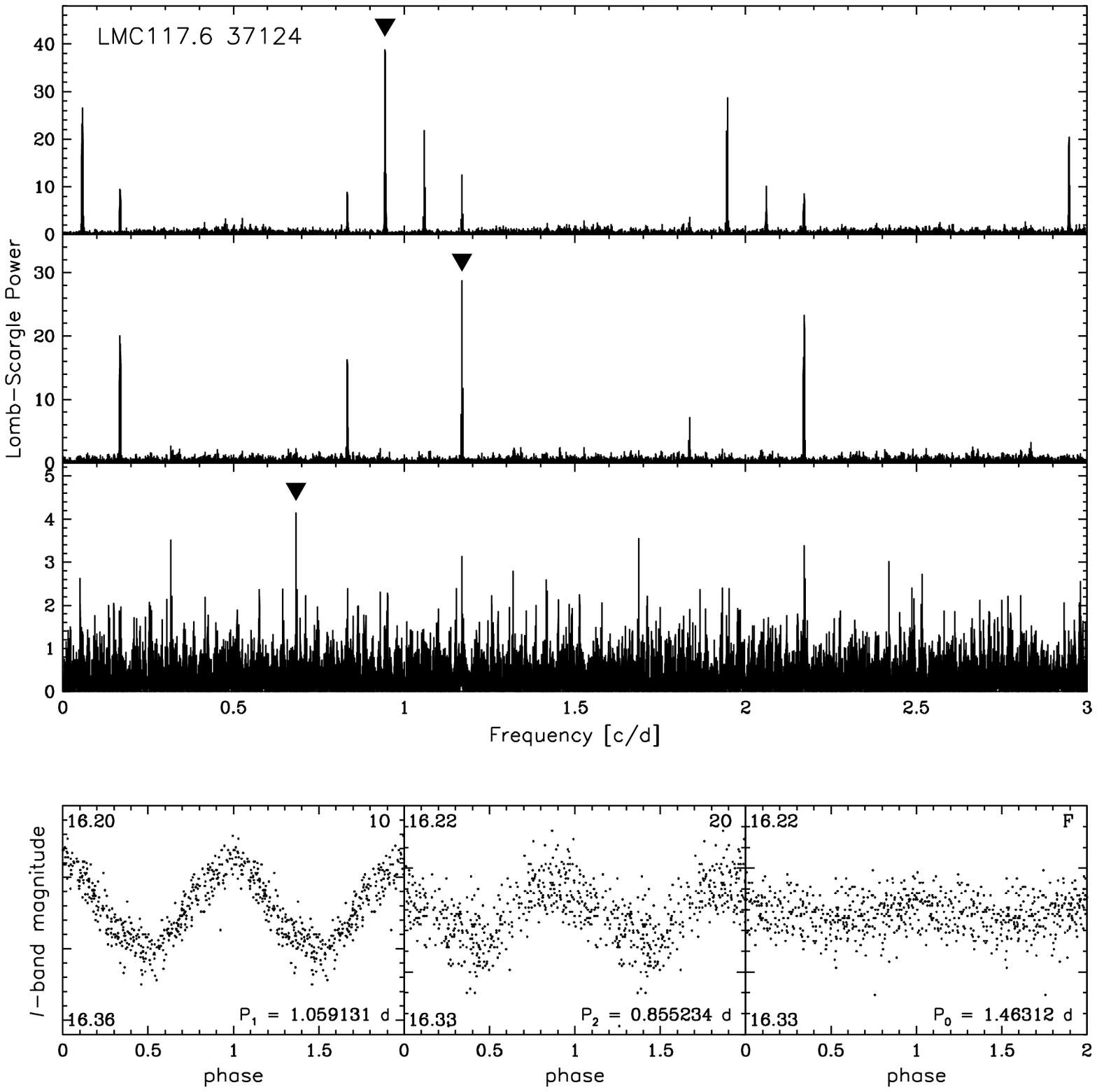}
\FigCap{Power spectra and light curves of F/1O/2O triple-mode Cepheid LMC117.6
37124. The description of the diagrams is the same as in Fig.~1.}
\end{figure}

A special attention was devoted to the selection of triple-mode
Cepheids. We checked all double mode Cepheids in our list for additional
variability. First, we prewhitened the light curves with two pulsational
periods and detrended the data by fitting and subtracting splines. Then, we
checked the residual data of every double-mode Cepheid, looking for
independent periodicities which may correspond to additional radial mode of
pulsation. All the stars with proper position in the Petersen diagram (a
plot of period ratios \vs longer periods) were visually inspected and five
likely candidates for triple-mode Cepheids were selected. All these objects
were found in the list of 1O/2O double-mode Cepheids. For each of them the
first overtone mode dominates in the power spectrum. For two of these
variables the period ratios identify the new periodicities as fundamental
modes of radial pulsations, while three remaining objects are 1O/2O/3O
triple-mode Cepheids. From the latter group only one object -- LMC179.4
33340 -- is a new identification; two other stars were detected by Moskalik
\etal (2004) in the OGLE-II dataset.

\begin{figure}[htb]
\includegraphics[width=12.5cm, bb=40 225 565 745]{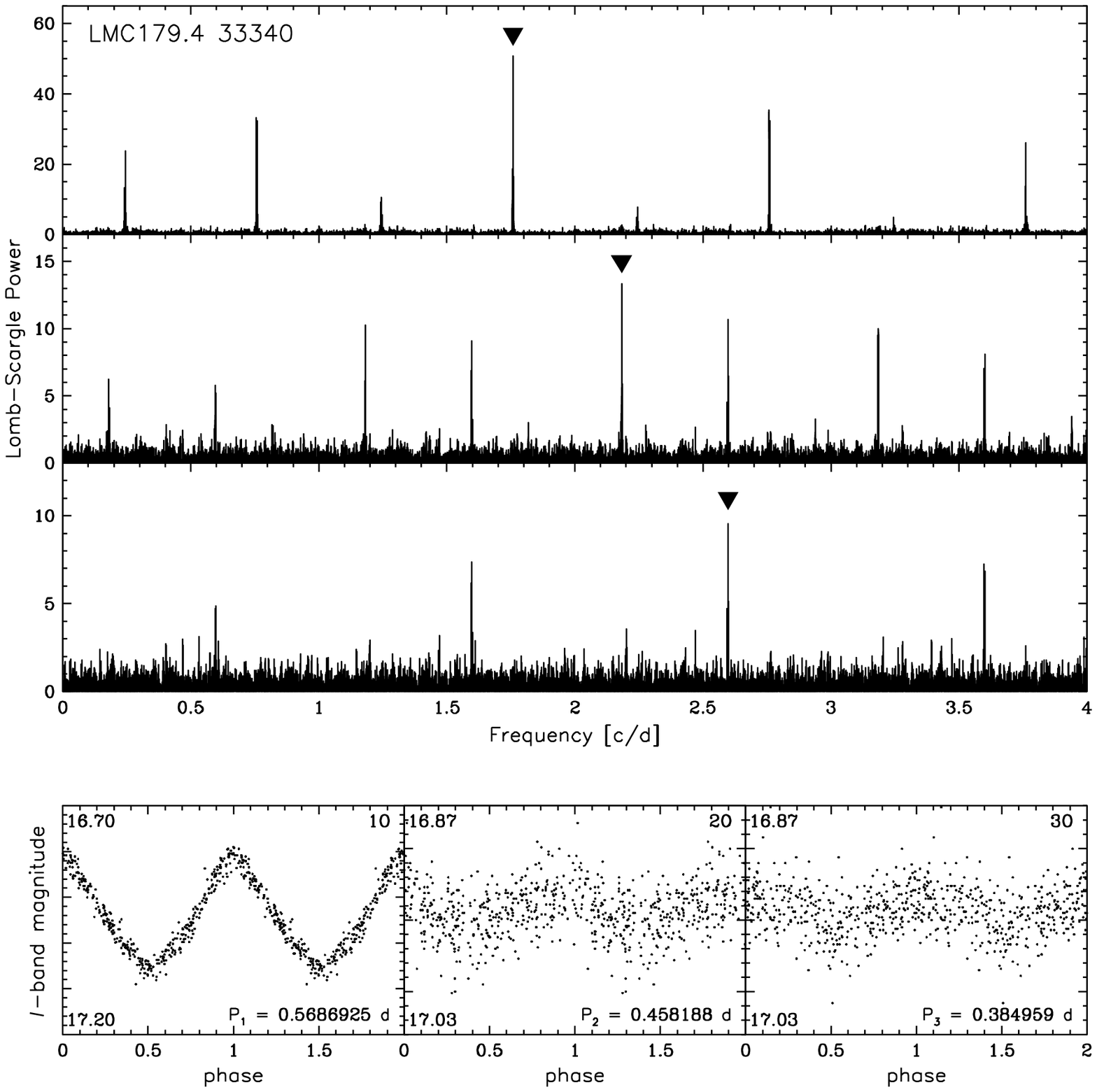}
\FigCap{Power spectra and light curves of 1O/2O/3O triple-mode Cepheid 
LMC179.4 33340. The description of the diagrams is the same as in Fig.~1.}
\end{figure}

\begin{figure}[p]
\psfig{figure=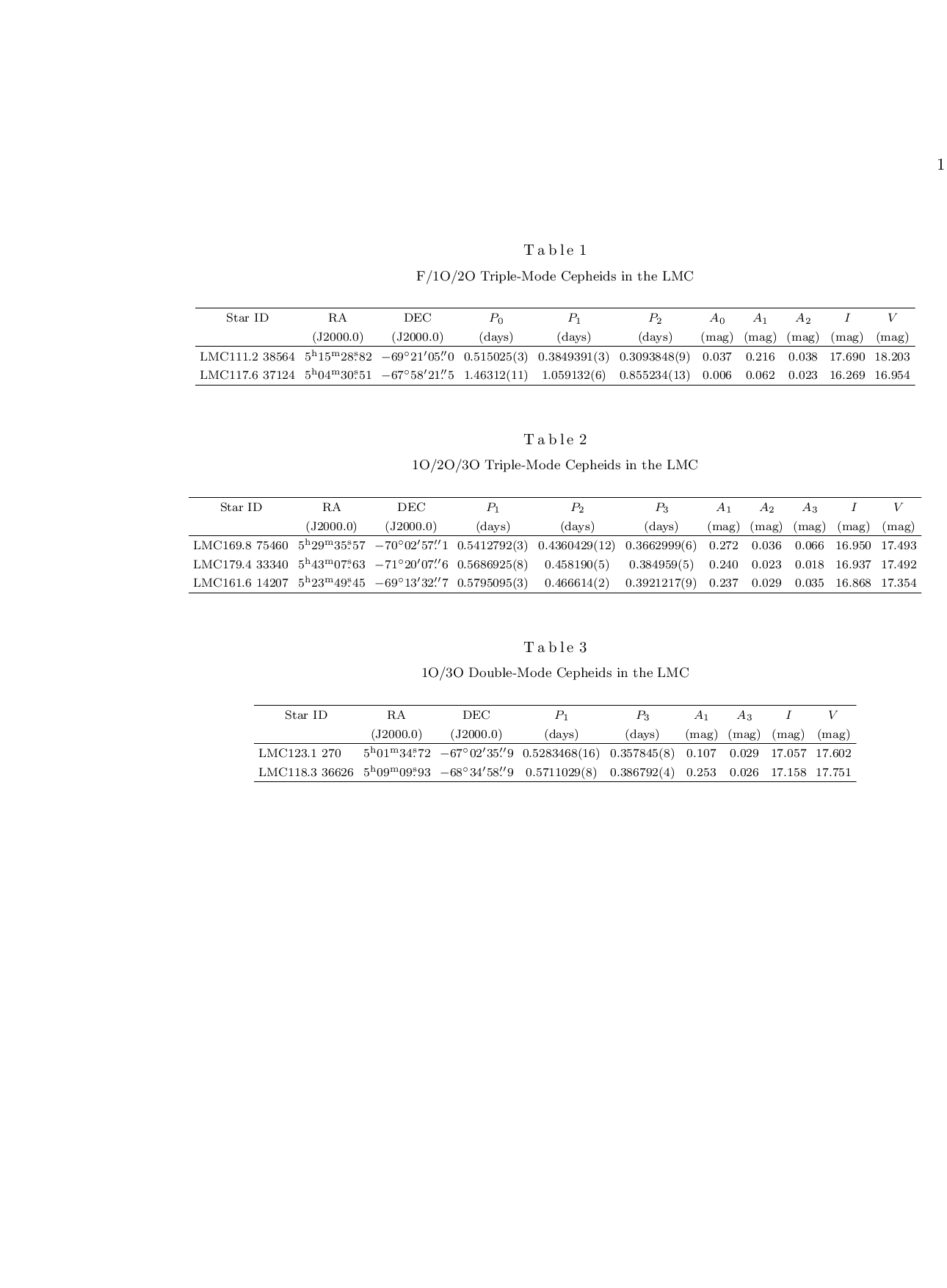,bbllx=0pt,bblly=0pt,bburx=670pt,bbury=685pt,angle=90,clip=}
\end{figure}

Tables 1 and 2 summarize the main observational properties of the
triple-mode Cepheids in the LMC. Amplitudes of the modes were derived as
differences between the maximum and minimum values of the functions fitted
to the light curves. Periodograms and light curves of newly detected stars
are shown in Figs.~1--3. In each figure the upper panel shows Lomb-Scargle
power spectrum obtained for the original data, the next panel contains
periodogram after removing the primary periodicity (it is always the first
overtone mode) and the lower spectrum is obtained with data after
prewhitening with the primary and secondary frequencies and their linear
combinations. Three panels below the spectra show the light curves folded
with the primary, secondary and tertiary periods but, in each case, after
subtracting the other two modes and their harmonics. Photometry and finding
charts of the stars presented in this paper can be found in Soszy{\'n}ski
\etal (2008).

\begin{figure}[htb]
\includegraphics[width=12.5cm, bb=40 330 565 745]{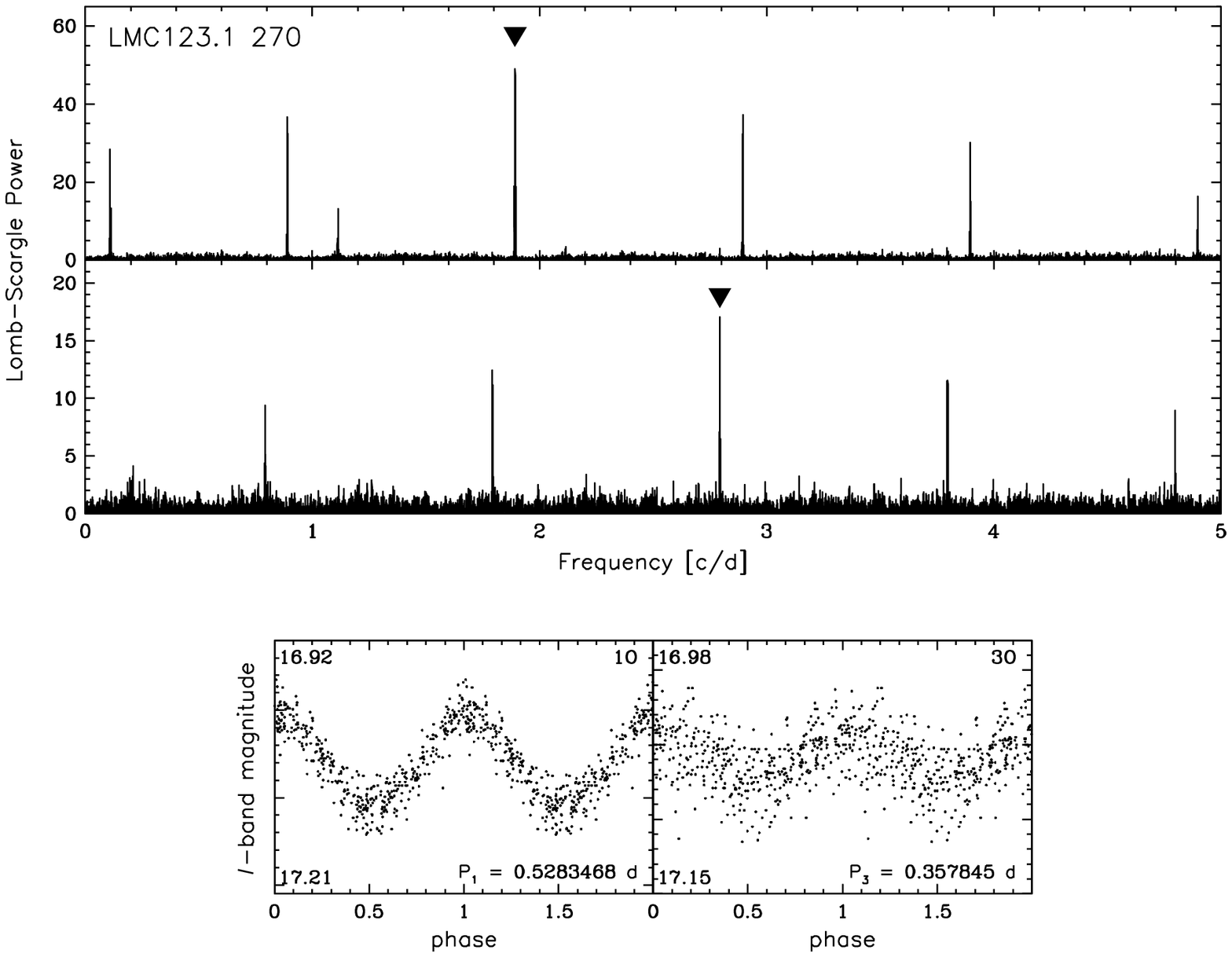}
\FigCap{Power spectra and light curves of 1O/3O double-mode Cepheid
LMC123.1 270. Two {\it upper panels} show power spectra for the original
data and after subtracting the primary periodicity, respectively. {\it
Bottom panels} display light curves folded with the primary and secondary
periodicities after prewhitening with the other period.}
\end{figure}

Then, we inspected the double-periodic Cepheids with unusual period
ratios. We paid attention for two objects with periods and period ratios
almost the same as for the first and the third overtones in the triple-mode
Cepheids, \ie $P_3/P_1=0.677$. Their position in the Petersen diagram was
practically the same as for $P_1$ and $P_3$ periods of triple mode
pulsators, but we detected no significant frequency corresponding to the
second overtone. We recognize these two stars as a new class of double-mode
Cepheids: 1O/3O pulsators. Table~3 lists the basic parameters of these
stars, while Figs.~4 and 5 show their power spectra and light curves. It is
worth mentioning that for LMC123.1 270 we detected very weak periodic
signal at 0.311617 days, what may be a signature of the fourth overtone
mode of pulsation.

\Section{Discussion}
The advantage of studying variable stars in the Magellanic Clouds is that
they can be easily placed in the period--luminosity (PL) diagrams, what
makes the variability type classification easier. All the stars presented
in this paper have the primary periods corresponding to the pulsation in
the first overtone and they match the PL sequence of classical Cepheids in
the {\it V} and {\it I}-band domains as well as in the $\log{P}$--$W_I$
plane, where $W_I=I-1.55(V-I)$ is the reddening-free Wesenheit index.

\begin{figure}[htb]
\includegraphics[width=12.5cm, bb=40 330 565 745]{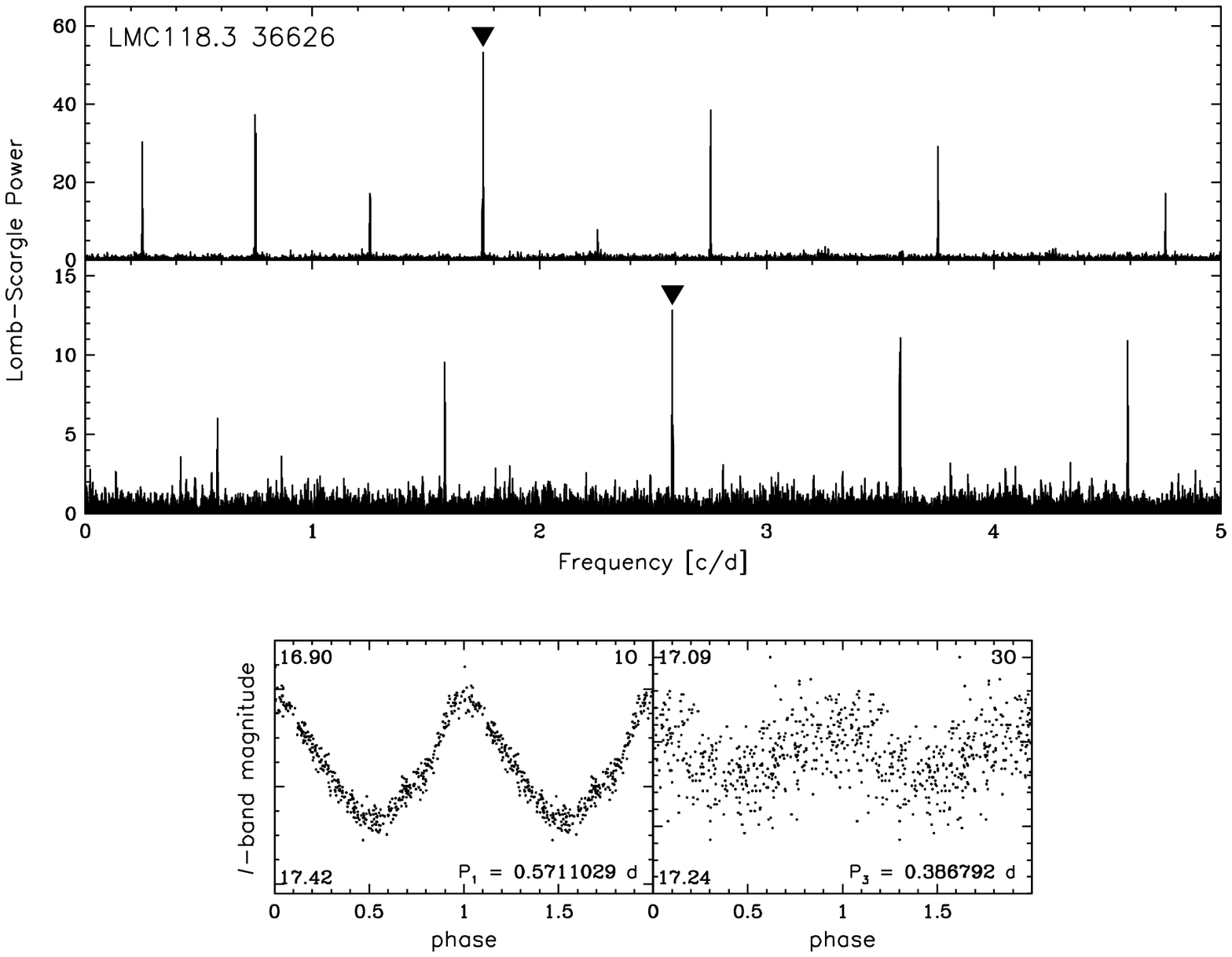}
\FigCap{Power spectra and light curves of 1O/3O double-mode Cepheid LMC118.3
36626. The description of the diagrams is the same as in Fig.~4.}
\end{figure}

The dominant periods and period ratios point that LMC111.2 38564 is of the
same class of variable stars as Galactic AC~And and V823~Cas, \ie pulsator
with the three lowest radial modes excited. Since LMC111.2 38564 is a
classical Cepheid, it should stop the discussion about the status of the
Galactic triple mode pulsators. In the literature AC~And and/or V823~Cas
were classified as RR~Lyr stars (Fitch and Szeidl 1976, Pe{\~n}a \etal
2006), analogs of $\delta$~Sct stars (Kov{\'a}cs and Buchler 1994), or as a
completely new class of variable stars (Rodr{\'i}guez 2002). In fact, the
fundamental-mode periods of these stars fall in the gap between classical
Cepheids and High Amplitude $\delta$~Sct (HADS) stars. However, our sample
of pulsating stars in the LMC shows no such a gap for the first-overtone
variables. Thus, it is rather a question of convention which period will be
chosen to distinguish between HADS and classical Cepheids. In the OGLE-III
Catalog of Variable Stars we adopted the limiting period equal to
$P_1=0.24$~days for the first overtone mode, what keeps our triple-mode
variables (together with AC~And and V823~Cas) on the Cepheid side.

LMC117.6 37124 is the candidate for ``long period'' triple-mode Cepheid
with the three lowest modes excited, but the third (fundamental mode)
periodicity is of marginal significance. The multiperiodicity of this star
requires confirmation with more accurate photometric data. We emphasize
that among our double mode Cepheids we detected more stars with very weak
additional periodic signal which may correspond to the radial mode of
pulsation. This suggests that asteroseismology of classical Cepheids would
yield new interesting information from dedicated precise photometric
surveys.

Three Cepheids with 1O, 2O and 3O modes excited constitute very homogeneous
group, with similar periods, luminosities and shapes of the light
curves. 1O/3O double-mode Cepheids seem to belong to the same class,
because they occupy the same, very narrow, region in the Petersen
diagram. The only difference between these two groups is the lack or
presence of the second overtone mode.
\vspace*{4pt}

Moskalik and Dziembowski (2005) analyzed two triple-mode Cepheids in the
LMC comparing their observed periods with those predicted from the
pulsational models. They concluded that both triple-mode Cepheids must be
on the first crossing of the instability strip. This result implies
considerable rate of period change expected for these stars, because the
evolution on the Hertzsprung gap is very fast.
\vspace*{4pt}

We made an attempt to measure the rate of period change in all stars
presented in this paper. To increase the time baseline of the photometry we
merged the OGLE photometry with publicly available data from the MACHO
project\footnote{{\it
http://wwwmacho.mcmaster.ca/Data/MachoData.html}}. After removing outlying
points we scaled amplitudes and shifted zero points of the MACHO $R_M$
photometric measurements to obtain the same values as for the OGLE {\it
I}-band data. Then, the two datasets were merged resulting in light curves
covering time span of more than 16 years (from 1992 to 2008).
\vspace*{4pt}

The search for secular period changes was performed with two methods. In
the first method, we folded data of each star using trial periods $P$ and
period change rates $\dot{P}$ and we fitted the Fourier series with the
same number of harmonics as in the period searching procedure (minimizing
the $\chi^2$ per degree of freedom). Then, using simplex method, we found
the values of $P$ and $\dot{P}$ that minimize the scatter around the fit to
the folded light curves. In the second method, we constructed $O-C$
diagrams for each star. We used an iterative procedure based on the
measurements of the phase shifts of the fitted mean light curve relative to
the chunks of data consisting of at least 10 observing points from the same
observing seasons. Then, the parabola was fitted to the $O-C$ diagram what
gave us the ephemeris with quadratic component. Details of both algorithms
will be described by Poleski (2008, in preparation).
\vspace*{4pt}

We tested only the first overtone periods, because the other modes have
much lower amplitudes and consequently the $\dot{P}$ are determined with
much larger errors. Both methods produced very similar period change rates
for each star, what ensured us that the procedures are reliable. Only for
two Cepheids we obtained the $\dot{P}$ significantly different from
zero. In both cases we measured the decrease of periods: for 1O/2O/3O
triple mode Cepheid LMC179.4 33340 we found $\dot{P}/P=-2.7\pm0.3\
\mathrm{Myr}^{-1}$ and for 1O/3O double-mode LMC123.1 270:
$\dot{P}/P=-1.5\pm0.5\ \mathrm{Myr}^{-1}$. These values are in clear
disagreement with the period change rates predicted by Moskalik and
Dziembowski (2005) for the first crossing of the instability
strip. However, these measurements disagree also with the hypothesis that
these stars are in the second crossing (if it is possible at all for stars
of that type), because the period changes are too fast. We conclude that
the observed changes of periods have probably other than evolutionary
reason.

\Acknow{We want to acknowledge Prof.~W.~Dziembowski for critical reading of
the manuscript and useful comments. This paper was partially supported by
the Polish MNiSW grants: NN203293533 to IS and N20303032/4275 to AU and by
the Foundation for Polish Science through the Homing Program.}

\end{document}